\documentclass{eptcs}
\usepackage[usenames,dvipsnames,svgnames]{xcolor}
\usepackage{url}

\usepackage{amsmath,amsopn,amssymb,amsthm}
\usepackage{mathtools}

\usepackage[caption=false]{subfig}
\usepackage{endnotes,microtype,xspace,graphicx,fancyvrb,multirow}

\usepackage{bold-extra}

\usepackage{cite}

\usepackage{array,underscore,relsize}
\usepackage[T1]{fontenc}
\usepackage{enumitem}
\usepackage{listings}
\usepackage{aliascnt}
\usepackage[linesnumbered, noend]{algorithm2e}
\usepackage{semantic}

\usepackage{stmaryrd}
\usepackage{verbatim}
\usepackage{makecell}

\usepackage{pifont}
\usepackage{booktabs}
\newcommand{\cmark}{\ding{51}}
\newcommand{\xmark}{\ding{55}}

\usepackage{tikz}


\fvset{ %
  fontsize=\scriptsize, %
  xleftmargin=8pt, %
  numbers=left, %
  numbersep=5pt
}

\input{code/fmt}

\def\Snospace~{\S{}}

\newif\ifnotes
\notestrue
\newcommand{\newnote}[2]{%
  \expandafter\newcommand\csname #1\endcsname[1]{%
    \ifnotes{%
      \textcolor{#2}{#1: ##1}%
    }\fi%
  }%
}
\newnote{BH}{OliveGreen}
\newnote{QZ}{blue}
\newnote{DAH}{orange}
\newnote{ALL}{red}

\input{glyphtounicode}
\newcommand{\sys}{\mbox{\textsc{Pequod}}\xspace}
\newcommand{\assign}{\mathbin{:=}}

\newcommand{\false}{\mathsf{False}}

\newcommand{\true}{\mathsf{True}}

\newcommand{\cc}[1]{\mbox{\smaller[0.5]\texttt{#1}}}
\newcommand{\bcc}[1]{\smaller[1.5]{\textbf{#1}}}

\newcommand{\ccall}[3]{#1 := #2(#3)}
\newcommand{\cif}[3]{\textsf{if}~#1~\textsf{then}~#2~\textsf{else}~#3~\textsf{fi}}
\newcommand{\com}{\textsf{Com}}
\newcommand{\pcom}{\textsf{PCom}}
\newcommand{\cskip}{\textsf{skip}}

\newcommand{\LTable}{\textbf{LTable}}

\newcommand{\triple}[3]{\{#1\}~#2~\{#3\}}
\newcommand{\judgement}[4]{T, #1 \vdash \triple{#2}{#3}{#4}}
\newcommand{\partrel}[3]{(#1, #2) = \textsf{part}(#3)}
\newcommand{\evaluation}[3]{\langle #1, #2 \rangle \Downarrow #3}

\newcommand{\Bounded}{\textsf{Bound}}

\newcommand{\ConstructCHC}{\textsf{ConstructCHC}}
\newcommand{\ConstructAux}{\textsf{ConstructAux}}
\newcommand{\Permute}{\textsf{Permute}}
\newcommand{\Partition}{\textsf{Partition}}
\newcommand{\Synthesize}{\textsf{Syn}}

\newcommand{\retvar}{\mathsf{ret}}

\gdef\therev{1c995b7}
\gdef\thedate{2019-06-18 15:44:56 -0400}

\title{Relational Verification via \\
Invariant-Guided Synchronization}

\ifdefined\DRAFT
\fi

\author{Qi Zhou %
\institute{Georgia Institute of Technology}
  \email{qzhou80@gatech.edu}
  \and David Heath %
\institute{Georgia Institute of Technology}
  \email{heath.davidanthony@gatech.edu}
  \and William Harris
\institute{Galois Inc.}
\email{wrharris@galois.com}
  }

\def\titlerunning{Relational Verification via Invariant-Guided Synchronization}
\def\authorrunning{Q. Zhou, D. Heath, \& W. Harris}

\begin{document}
\maketitle

% Abstract:
\begin{abstract}
  Relational properties describe relationships that hold over multiple
  executions of one or more programs, such as functional equivalence.
  Conventional approaches for automatically verifying such properties
  typically rely on syntax-based, heuristic strategies for finding
  \emph{synchronization points} among the input programs.
  These synchronization points are then annotated with appropriate
  relational invariants to complete the proof.
  However, when suboptimal synchronization points are chosen the
  required invariants can be complicated or even inexpressible in the
  target theory.

  In this work, we propose a novel approach to verifying relational
  properties.
  This approach searches for synchronization points and synthesizes
  relational invariants \emph{simultaneously}.
  Specifically, the approach uses synthesized invariants as a guide
  for finding proper synchronization points that lead to a complete
  proof.
  % evaluation:
  We implemented our approach as a tool named \sys,
  which targets Java Virtual Machine (JVM) bytecode.
  We evaluated \sys by using it to solve verification challenges drawn
  from the from the research literature and by verifying properties of
  student-submitted solutions to online challenge problems.
  The results show that \sys solve verification problems that cannot
  be addressed by current techniques.
\end{abstract}

\section{Introduction}
\label{sec:introduction}
\emph{Relational} properties characterize multiple executions of one
or more programs~\cite{sousa16}.
One example of such a property is that a particular program $f$ over
integers is monotonic;
i.e.,
\[ \forall x, y.\ x > y \Rightarrow f(x) > f(y) \] This property is
relational because it is defined over two arbitrary inputs of $f$
(named $x$ and $y$, respectively).
Relational properties can express important problems, such as the
equivalence of two programs or the information-flow security of a
single program.
Therefore, a tool that could automatically verify relational
properties would be highly valuable both to program developers and
users.

% Existing approaches
Substantial effort has been directed toward constructing relational
\emph{verifiers}, which attempt to prove that given programs satisfy a
given relational property.
One effective approach attempts to synthesize proofs in Cartesian
Hoare Logic~\cite{sousa16,Lauren18}, which extends Hoare Logic from individual
programs to tuples of programs.
Such approaches consider the execution of input programs
simultaneously, which enables the construction of relational
invariants that describe relationships across the programs.
By examining the programs together, a verifier can potentially find
simpler invariants than if it had attempted to summarize each program and
then compared the summaries.

% why it is hard
% BH: fix from here
However, synthesizing proofs in such a system adds a critical new
dimension to a verifier's design.
In particular, a verifier must choose pairs of control locations to
relate, in addition to synthesizing sufficiently strong invariants
that relate the programs' data when they reach related locations.
Such pairs of locations are referred to as \emph{synchronization
  points}~\cite{Lauren18}.
Intuitively, certain synchronization points can be annotated with
simple relational invariants to form proofs because the variables at
the related points maintain similar data.
Conversely, non-ideal synchronization points may relate locations for
which sufficient invariants can be expressed using only complex
formulas, or even formulas expressible only in complex theories and
logics.

% loop is harder:
Selecting synchronization points is particularly difficult when a
verifier must prove a relational property over programs with loops or
recursive procedures.
Specifically, finding an ideal set of synchronization points may
require the verifier to consider different numbers of iterations for
different loops.
As an example, a suitable synchronization strategy might be to model
two iterations of a loop in one program for every one iteration of a
loop in a second program.
This highlights that no straightforward solution, such as modeling
each loop exactly once, is effective in general.

For the reasons given above, it is clear that finding effective
strategies for selecting synchronization points is an important and
difficult problem.
The effectiveness of a selecting synchronization points depends on
the data relationships in the programs and the property.
However, existing approaches~\cite{sousa16,Lauren18} have relied on
syntax-driven, heuristic strategies that first find synchronization
points and then attempt to annotate the points with relational
invariants to complete the proof.
The effectiveness of these strategies usually heavily depends on the
programs to which they are applied.

% our approach:
In this paper, we propose a general, automatic technique for
synthesizing proofs of relational properties.
The key feature of our approach is that it searches the spaces of
potential synchronization points and their relational invariants
\emph{simultaneously}.
% how it works
Our approach iteratively operates on a sequence of bounded
under-approximations of input programs;
in each bounded under-approximation, each recursive procedure call is
only allowed to execute a bounded number of times.
In each iteration, our approach attempts to generate a set of proofs
that the bounded programs satisfy the given relational property under
\emph{all} possible relevant choices of synchronization points.
Our approach synthesizes this set of proofs by solving a single system
of \emph{Constrained Horn Clauses}.
Then, our approach attempts to find \emph{some} proof of the
correctness of the bounded under-approximations that can be
generalized to form a proof for the original, unbounded programs.
If a valid proof is found, then the verifier has validated the given
relational property.
Otherwise, our approach continues by considering larger
under-approximations of the input programs.

% evaluation highlights:
We have implemented our approach as an executable tool, named \sys.
\sys targets Java Virtual Machine (JVM) bytecode and has been
evaluated on $33$ benchmarks, consisting of verification challenge
problems and student solutions submitted to online coding platforms.
Our evaluation indicates that, in a significant set of practical cases,
\sys can efficiently verify relational properties beyond the scope of
existing techniques.

% Paper outline.
The rest of this paper is organized as follows.
\autoref{sec:overview} provides an informal overview of our proof
system and of \sys, by example.
\autoref{sec:background} reviews the technical foundations for our
work, and \autoref{sec:approach} presents the proof system and \sys in
detail.
\autoref{sec:evaluation} presents an empirical evaluation of \sys.
\autoref{sec:related-work} concludes by comparing our contribution to
related work.

\section{Overview}\label{sec:overview}
In this section, we illustrate our approach by example.
We first introduce a pair of programs that compute the same function.
We formalize a relational property that these two programs are
equivalent as an extended Hoare Logic Triple.
Next, we describe how \sys~finds a proof of this triple in
\autoref{sec:example-proof}.
%
% running example:
\begin{figure}[t]
  \begin{minipage}[t]{0.45\linewidth}
    \input{code/add-0.java}
  \end{minipage}
  \begin{minipage}[t]{0.45\linewidth}
    \input{code/add-1.java}
  \end{minipage}
  \caption{%
    \cc{tri0} and \cc{tri1}: equivalent programs that, given
    integer $n$, compute the $n$th triangle number.%
  }\label{fig:running-ex-code}
\end{figure}

% walk through tri0:
\autoref{fig:running-ex-code} contains two programs, named
\cc{tri0} and \cc{tri1}, that each compute the $n$th
\emph{triangle number}: i.e., the sum of all natural numbers up to and
including $n$.
$\cc{tri0}$ computes this value by direct recursion while
$\cc{tri1}$ makes use of an auxilliary procedure,
$\cc{tri1Aux}$, which maintains an accumulator.
Despite these differences, these two programs compute the same
function.

% declare the relational property we want to prove
To verify this equivalence, we can construct a relational property that shows
that given equal parameters $n$, \cc{tri0} and \cc{tri1}
compute the same output.
This property can be represented as a Hoare Logic Triple over a
\emph{product} command:
\[
\triple{\cc{n}_0 = \cc{n}_1}{\cc{tri0} \times \cc{tri1}}{\retvar_0 =
\retvar_1}\]
The product command $\cc{tri0} \times \cc{tri1}$ can be
understood as the command that executes $\cc{tri0}$ and
$\cc{tri1}$ simultaneously.
A detailed explanation of product commands is given in
\autoref{sec:proof-system}.
We annotate variables with subscripts $0$ or $1$ to indicate which program they
model.
Variables $\retvar_0$ and  $\retvar_1$ are used to model the output of
the respective programs.
We refer to the proposition $\cc{n}_0 = \cc{n}_1$ as the
pre-condition, and the proposition $\retvar_0 = \retvar_1$ as the
post-condition.
The triple above states that if $\cc{n}_0 = \cc{n}_1$ and both
$\cc{tri0}$ and $\cc{tri1}$ are executed, then both programs
will return the same value.
A proof of this triple would prove the equivalence of the two
programs.

\subsection{Proving Equivalence Automatically}\label{sec:example-proof}

\sys~proves this example Hoare Triple in three steps.
First, \sys~constructs bounded versions of $\cc{tri0}$ and $\cc{tri1}$
that respect an upper bound on the allowed number of recursive
procedure calls. In this example, we set this upper bound to three.
%
% code for running examples:
\begin{figure}[t]
  \begin{minipage}[t]{0.45\linewidth}
    \input{code/addB0.java}
  \end{minipage}
  \begin{minipage}[t]{0.45\linewidth}
    \input{code/addB1.java}
  \end{minipage}
  \caption{%
    $\cc{tri0}_0$ and $\cc{tri1}_0$ are under-approximations
    of the input programs.
  }\label{fig:bounded-code}
\end{figure}
\autoref{fig:bounded-code} lists the bounded programs
$\cc{tri0}_0$ and $\cc{tri1}_0$.
These two program are bounded because each has finitely many execution
paths.
These two programs under-approximate $\cc{tri0}$ and
$\cc{tri1}$ respectively, because their execution paths are a
subset of the execution paths in the original programs.
$\cc{tri0}_2$ and $\cc{tri1Aux}_2$ are incomplete because
they do not have else branches in their conditional statement.
These branches are assumed to be unreachable in the current
under-approximation.

Second, \sys~tries to synthesize a set of proofs for a corresponding
Hoare Triple over these bounded programs:
$\triple{\cc{n}_0 = \cc{n}_1}{\cc{tri0}_0 \times \cc{tri1}_0}{\retvar_0 = \retvar_1}$.
The key idea is that \sys~will find proofs for \emph{all} possible
orders of modeling the execution of $\cc{tri0}_0$ and $\cc{tri1}_0$.
The resulting proofs represent all possible choices of synchronization
points of the bounded programs.
For example, in a subset of the bounded proofs, \sys~arrives at the
following intermediate goal:
\[\triple{\cc{n}_0 = \cc{x}_1}{\cc{tri0}_0 \times \cc{tri1Aux}_0}{\retvar_0 + \cc{acc}_1 = \retvar_1 }\]
\sys~continues the proof by either stepping in $\cc{tri0}_0$ or by
stepping in $\cc{tri1}_0$.
`Stepping' through the product program corresponds to applying
particular proof rules that result in new Hoare Triple goals.
In other words, \sys~proves the triple over
$\cc{tri0}_0 \times \cc{tri1Aux}_0$ by proving a series of Hoare
Triples, which we refer to as a proof path, with proper invariants.

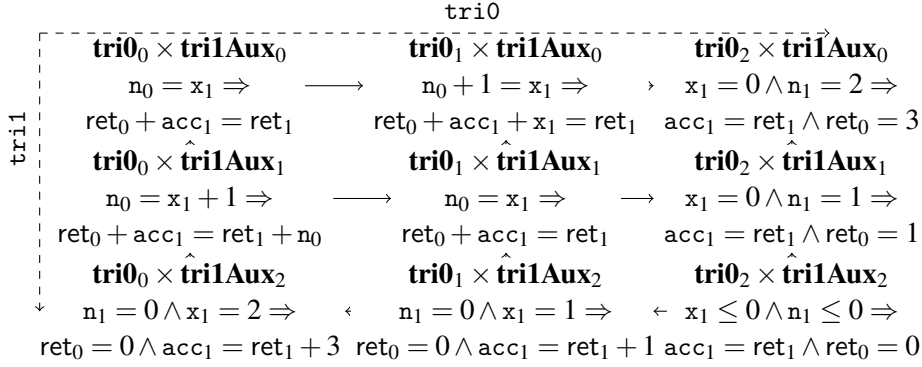
\begin{figure}[t]
  \centering
  \begin{tikzpicture}
  \path[dashed, ->]
    (5.25, 4) node(tri0) {\cc{tri0}}
    (-0.5, 3.7) edge node {} (10, 3.7)
    ;
  \path[dashed, ->]
    (-0.8, 2.3) node[rotate=90](tri1) {\cc{tri1}}
    (-0.5, 3.7) edge node {} (-0.5, 0)
    ;
  \path[->]
    (1.5,3) node(n00) [align=center] {%
      $\bcc{tri0}_0 \times \bcc{tri1Aux}_0$\\
      $\cc{n}_0 = \cc{x}_1 \Rightarrow$\\
      $\retvar_0 + \cc{acc}_1 = \retvar_1$
    }

    (5.7,3) node(n01) [align=center] {%
      $\bcc{tri0}_1 \times \bcc{tri1Aux}_0$\\
      $\cc{n}_0 + 1 = \cc{x}_1 \Rightarrow$\\
      $\retvar_0 + \cc{acc}_1 + \cc{x}_1 = \retvar_1$
    }

    (9.5,3) node(n02) [align=center] {%
      $\bcc{tri0}_2 \times \bcc{tri1Aux}_0$\\
      $\cc{x}_1 = 0 \land \cc{n}_1 = 2 \Rightarrow$\\
      $\cc{acc}_1 = \retvar_1 \land \retvar_0 = 3$
    }

    (1.5,1.5) node(n10) [align=center] {%
      $\bcc{tri0}_0 \times \bcc{tri1Aux}_1$\\
      $\cc{n}_0 = \cc{x}_1+1 \Rightarrow$\\
      $\retvar_0 + \cc{acc}_1 = \retvar_1 + \cc{n}_0$
    }

    (5.7,1.5) node(n11) [align=center] {%
      $\bcc{tri0}_1 \times \bcc{tri1Aux}_1$\\
      $\cc{n}_0 = \cc{x}_1 \Rightarrow$\\
      $\retvar_0 + \cc{acc}_1 = \retvar_1$
    }

    (9.5,1.5) node(n12) [align=center] {%
      $\bcc{tri0}_2 \times \bcc{tri1Aux}_1$\\
      $\cc{x}_1 = 0 \land \cc{n}_1 = 1 \Rightarrow$\\
      $\cc{acc}_1 = \retvar_1 \land \retvar_0 = 1$
    }

    (1.5,0) node(n20) [align=center] {%
      $\bcc{tri0}_0 \times \bcc{tri1Aux}_2$\\
      $\cc{n}_1 = 0 \land \cc{x}_1 = 2 \Rightarrow$\\
      $\retvar_0 = 0 \land \cc{acc}_1 = \retvar_1 + 3$
    }

    (5.7,0) node(n21) [align=center] {%
      $\bcc{tri0}_1 \times \bcc{tri1Aux}_2$\\
      $\cc{n}_1 = 0 \land \cc{x}_1 = 1 \Rightarrow$\\
      $\retvar_0 = 0 \land \cc{acc}_1 = \retvar_1 + 1$
    }

    (9.5,0) node(n22) [align=center] {%
      $\bcc{tri0}_2 \times \bcc{tri1Aux}_2$\\
      $\cc{x}_1 \leq 0 \land \cc{n}_1 \leq 0 \Rightarrow$\\
      $\cc{acc}_1 = \retvar_1 \land \retvar_0 = 0$
    }

    % right edges
    (n00) edge node {} (n01)
    (n01) edge node {} (n02)
    (n10) edge node {} (n11)
    (n11) edge node {} (n12)
    (n20) edge node {} (n21)
    (n21) edge node {} (n22)

    % down edges
    (n00) edge node {} (n10)
    (n10) edge node {} (n20)
    (n01) edge node {} (n11)
    (n11) edge node {} (n21)
    (n02) edge node {} (n12)
    (n12) edge node {} (n22)
    ;
\end{tikzpicture}
  \caption{%
    Intermediate products that appear when proving the
    bounded example program. Each product is depicted with a pre and
    post invariant that leads to a proof.
  }\label{fig:subset-proof}
\end{figure}

The fact that $\cc{tri0}_0$ and $\cc{tri1}_0$ are bounded commands
implies that there are finitely many possible proof paths that can be
used to prove this bounded goal.
\autoref{fig:subset-proof} depicts all possible proof paths for a
partial proof of this Hoare Triple.
The depicted proof is partial because we omit proof goals
corresponding to the programs' false branches, for clarity.
The upper-leftmost node is the Hoare Triple that \sys~must prove.
Every proof path over the true branches eventually reaches the the
product command $\cc{tri0}_2 \times \cc{tri1Aux}_2$ (since
$\cc{tri0}_2$ and $\cc{tri1Aux}_2$ are under-approximations that allow
no further recursion).
\sys~encodes all possible proof paths into a single set of Constrained Horn
Clauses (CHCs), and uses known techniques for solving this system to
synthesize proper invariants.
In short, \sys~uses CHCs to synthesize a set of proofs for all
possible proof paths of the bounded program. The method for converting
an input Hoare Triple into a CHC system is described in
\autoref{sec:solve}.

Third, \sys~attempts to prove the original, unbounded problem by
searching for a bounded proof that can be generalized.
A suitable approach for solving this example is to consider the
synchronization point $\cc{tri0} \times \cc{tri1Aux}$, since the data
at these two points is highly related.
\sys~finds this synchronization point automatically and annotates it
with appropriate invariants by searching the set of proofs for the
bounded programs, as depicted in \autoref{fig:subset-proof}.
\sys~begins its search from the top-left node: $\cc{tri0}_0 \times \cc{tri1Aux}_0$.
%$\triple{\true}{\cc{tri0}_0 \times \cc{tri1Aux}_0}{\cc{n}_0
%= \cc{x}_1 \Rightarrow \retvar_0 = \retvar_1 + \cc{acc}_1}$.
%
In order to find a generalizable proof path, \sys~must choose a proof
path that passes through the node $\cc{tri0}_1 \times \cc{tri1Aux}_1$.
%$\triple{\true}{\cc{tri0}_1 \times \cc{tri1Aux}_1}{\cc{n}_0
%= \cc{x}_1 \Rightarrow \retvar_0 = \retvar_1 + \cc{acc}_1}$.
%
Note that the pre-condition and post-condition of two Hoare Triples over these two product commands
are the same. 
Furthermore, the two product commands represent the same unbounded command,
$\cc{tri0} \times \cc{tri1Aux}$, from the original program.
Thus, \sys~can use the Hoare Triple over the first command as a hypothesis to prove the
second. The details of the proof rule that allows this reasoning is
given in \autoref{sec:proof-system}.
By choosing this proof path, \sys~has also decided to synchronize the
two procedures by executing each recursive procedure once.
Therefore, \sys~finds the proof and synchronization points simultaneously.

Not all proof paths can be generalized to form proofs for the original
programs.
In fact, any proof path that does not pass through the node $\cc{tri0}_1 \times
\cc{tri1Aux}_1$ will fail to generalize because \textbf{(1)} the
relational invariants along the path are inappropriate for use as a
hypothesis further down the path and \textbf{(2)}
the node $\cc{tri0}_2 \times \cc{tri1Aux}_2$ cannot be used in the
proof because it incompletely models the original programs.
When \sys cannot find a valid proof for the original programs, \sys
increases the bounding number and starts over.
The algorithm that generalizes bounded proofs is described in
\autoref{sec:choose}.

\section{Background}
\label{sec:background}
In this section, we present the technical background for our approach.
In \autoref{sec:target-language}, we formalize the imperative target
language.
In \autoref{sec:chc}, we introduce Constrained Horn Clause (CHC)
systems as a class of logic-programming problems.

\subsection{Target Language}
\label{sec:target-language}
In this section, we give the formal definition of the target language:
an imperative language with conditionals and (possibly recursive)
procedures.
In order to define a program, we first give the definition of a
command.
In the following, we use the metavariable $x$ to represent program
variables, $e$ to represent program expressions, and $N$ to represent
procedure names.
The space of commands $\com$ is defined inductively, as follows:
\begin{align*}
  \com ::=
  \cskip~|~x := e
        ~|~\cif{e}{\com}{\com}
        ~|~\ccall{\vec{x}}{N}{\vec{x}}
        ~|~\com~;~\com
\end{align*}

That is, a command is either a $\cskip$ command, an assignment,
a conditional, a procedure call, or a sequence of other commands.

\begin{figure}[t]
  \input{fig/eval-rules}
  \caption{%
    Operational semantics of programs in \com.
  }\label{fig:semantics}
\end{figure}
The semantics of this language is defined in terms of program states
that the commands manipulate.
A program state, $\sigma$, is a map from variables to values:
$\sigma : x \rightarrow v$.
\autoref{fig:semantics} formalizes the semantics of commands by
relating states before and after executing the command.
A $\cskip$ command leaves the state unchanged (\textsc{E-Skip}).
An assignment updates a program variable $x$ to an given value
$e$ (\textsc{E-Assign}).
A conditional first evaluates the condition expression $e$, and if the
result of evaluation result is the symbol $\true$, then its evaluation
is based on the first command (\textsc{E-IfTrue}); otherwise, the
evaluation is based on the second command (\textsc{E-IfFalse}).

In order to support procedures, we also define a space of lookup
tables $\LTable$.
A lookup table maps the names of procedures to a
tuple: the parameters $\vec p$, the body of the procedure $\com$, and
the output variables $\vec o$.
\[ \LTable : N \rightarrow (\vec p, \com, \vec o) \]
A procedure call applies a lookup table $T \in \LTable$ to a procedure name $N$,
constructs a program state $\sigma'$ by substituting the parameters
$\vec{i}$ by the arguments $\vec{x}$, evaluates the body over
$\sigma'$ to get $\sigma''$, and finally substitutes the return
variables $\vec{r}$ by the value of the output variables $\vec{o}$ in
$\sigma''$.

A program is a command paired with a lookup table.

\subsection{Constrained Horn Clauses}\label{sec:chc}
\sys~finds valid program invariants using an external Constrained Horn
Clause (CHC) solver.
CHCs are a class of constraint programming problems~\cite{bjorner13}.
Each CHC has the following form:
$\cc{chc} := \cc{head} \Leftarrow \cc{body}$.
The clause head is an uninterpreted predicate applied to a set of
variables.
The clause body is the conjunction of a logical formula together with
any number of uninterpreted predicates applied to variables.
A CHC system is a set of CHCs together with a distinguished
uninterpreted predicate, called the query.

A \emph{solution} of a CHC system is a map from
uninterpreted predicates to interpretations.
A \emph{valid} solution is one where \textbf{(1)} the interpretation
of the query is the constant function that returns the proposition
$\false$ and \textbf{(2)} replacing each uninterpreted predicate by
its interpretation (instantiated with the applied variables) results
in a set of valid implications.
%
%Constructing efficient, automatic CHC system solvers is an active area
%of research \DAH{\QZ{references}}.

\section{Technical Approach}\label{sec:approach}
In this section, we describe our technical approach in detail.
In \autoref{sec:proof-system}, we define the proof system that \sys
uses to prove relational properties 
In \autoref{sec:prove}, we describe a procedure for automatically
finding proofs in this system.
\subsection{Proof System}
\label{sec:proof-system}
We define a proof system that extends standard Hoare Logic with new rules
that can verify relational properties.
To present this system, we first define the concept of a product
command:
  $\pcom ::= \com \times \com$.
Informally, product commands are used to represent pairs of independent
program fragments whose execution we wish to consider simultaneously.
This intuition can be formalized by the following semantic rule:
\[
  \inference[\textsc{E-Prod}]{%
    \evaluation{c_0}{\sigma}{\sigma'}&
    \sigma \cap \tau = \varnothing\\
    \evaluation{c_1}{\tau}{\tau'}&
    \sigma' \cap \tau' = \varnothing\\
  }{%
    \evaluation{c_0 \times c_1}{\sigma \cup \tau}{\sigma' \cup \tau'}
  }
\]
Because members of a product command share no vocabulary, we can
reorder the members at will without changing the semantic meaning:
The product of two commands is commutative with respect to the program
semantics.
Additionally, we enrich the vocabulary of commands with one additional
constructor: $\llbracket\com\rrbracket$. This command essentially adds
a wrapper around the inner command. The semantic meaning of the
wrapped command is the same as the inner command; we merely add this
construction for the purposes of the proof rules.

Given these additional constructions, our proof system extends the
standard Hoare Logic naturally such that it respects products.
A relational invariant, $P$, is a first-order logical proposition that
contains the vocabulary of each program.
Judgments in the proof system take the following form:
\[\judgement{\Gamma}{P}{\pcom}{Q}\]
$P$ and $Q$ are relational invariants, where $P$ is the precondition
and $Q$ is the postcondition.
$T$ is a lookup table that contains mappings from procedure names to
procedure bodies (implemented as commands).
$\Gamma$ is a context, which is a set of Hoare Triples of the form $\triple{P}{\pcom}{Q}$.
$\Gamma$ is used as a set of hypotheses which can be used to complete
proofs of programs with recursion.

\begin{figure}[t]
  \input{fig/proof-rules}
  \caption{%
    Proof judgments for determining the validity of invariants over
    product command.
  }\label{fig:judgments}
\end{figure}

\autoref{fig:judgments} presents the proof rules.
Rules $\textsc{Skip}$ and $\textsc{Assign}$ simply model the semantics
associated with the respective command in the context of a product.
$\textsc{If}$ models the semantics of a conditional command.
A critical difference between this rule and the rule from
standard constructions is that the conditional is part of a product command. This
allows the prover to reason about both branches of a conditional
simultaneously with another program.

$\textsc{Call}$ models the semantics of a procedure call.
Suppose that the prover wishes to prove an assertion over a pair of
two commands: a call command $x \assign N(e)$ with an arbitrary
command $c$.
We use $N$ for the name of the called procedure, $c_0$ for the body of
the procedure, $\vec{p}$ for the vector of parameters, and $\vec{o}$
for the vector of return variables.
If the prover demonstrates that the
\emph{wrapped} command $\llbracket c_0 \rrbracket $ paired with $c$ satisfies pre-condition
$P$ and post-condition $R$, then
the call to $N$ paired with $c$ satisfies $Q$, 
given that the pre-condition $P[ \vec{p} \mapsto \vec{e} ]$ holds under an additional assumption.
In \autoref{fig:judgments}, the assumption
$\forall X'. (R[ X \mapsto X' ] \implies Q[ \vec{x} \mapsto \vec{o} ]$
means that $R$, which holds at the end of the called procedure,
entails the post-condition $Q$ in callee procedure, after substituting
the output variables $\vec{x}$
by the return variables $\vec{o}$.
$X'$ a copy of variables in callee that are different from the output variables.

$\textsc{Step}$ is used to step into the body of a procedure.
It allows
the prover to add the current goal as a hypothesis. The rule unwraps
the command while adding the goal to the hypothesis.
Later, the prover can use $\textsc{Assume}$, which states that a proof
goal can be satisfied if the goal is a hypothesis in the context
$\Gamma$.
$\textsc{Cons}$ is a typical component of a Hoare Logic system. It
states that we can weaken the pre-condition and strengthen the
post-condition.
As stated earlier, the semantics of commands are commutative with
respect to products. $\textsc{Comm}$ allows the prover to continue
the proof by applying $\textsc{Assign}$, $\textsc{If}$, and
$\textsc{Call}$ on either member of the product.
This rule is critical for relational reasoning. In practice, the
prover uses $\textsc{Comm}$ to select the order in which to model the
subcommands.
The example in \autoref{sec:example-proof} shows that choosing the
right order to apply $\textsc{Comm}$ results in simple invariants that
satisfy the proof.
Recall that a focus of our approach is finding appropriate
synchronization points of the program in conjunction with relational
invariants.
$\textsc{Part}$ is responsible for this reasoning.
$\textsc{Part}$ makes use of a procedure $\textsf{part}$.
Informally, $\textsf{part}$ allows the prover to partition a sequence
of commands into two subsequences by cutting a sequence at an
arbitrary point.
$\textsc{Part}$ decomposes a product command into two product
commands, and proves them sequentially.
The formal definition of $\textsf{part}$ is given in the extended paper.

One key observation of this proof system is that this system is
non-deterministic.
In particular, $\textsc{Part}$ allows the prover to subdivide the
input programs at will: By choosing different partitionings, the
prover is selecting synchronization points.
$\textsc{Comm}$ rule is also non-deterministic, and can be applied
anywhere in the proof.
Once a suitable ordering has been chosen by applying $\textsc{Comm}$
and $\textsc{Part}$, the prover can potentially construct simple
invariants that lead to a valid proof.
Hence, the difficulty of designing the automatic proof system is
determining how to use $\textsc{Comm}$ and $\textsc{Part}$.

\subsection{Verifying Relational Properties Automatically}\label{sec:prove}
Verifying a relational property of a product command, $pcd$,
is
reducible to deriving a relational Hoare Triple $\triple{P}{pcd}{Q}$
under a given the lookup table $T$ and an empty context $\Gamma$.
The relational property is modeled by the pre-condition $P$ and
the post-condition $Q$.
For example, the Hoare Triple given in \autoref{sec:overview}
describes a property that specifies $\cc{tri0}$ and $\cc{tri1}$ are equivalent.

\sys attempts to construct a proof of a relational Hoare Triple by
iteratively executing three steps:
First, \sys constructs a bounded product command $pcd'$ from the
original product command $pcd$ that respects a given bounding number
$n$.
$pcd'$ is an under-approximation of $pcd$ where each recursive procedure
executes at most $n$ times.
In \autoref{sec:bounded}, we describe how to construct $pcd'$ from $pcd$
and $n$.
Second, \sys generates a set of proofs for $pcd'$ in a corresponding
Hoare Triple.
Because $pcd'$ is bounded, \sys can attempt all proof paths by
exhaustively applying $\textsc{Comm}$ and $\textsc{Part}$.
\sys populates these proofs with appropriate intermediate invariants
using a system of Constrained Horn Clauses.
In \autoref{sec:solve}, we describe how to generate a set of proofs
for a bounded Hoare Triple.
In the third step, \sys attempts to generalize the work done in the
second step by finding a proof for the unbounded commands among the
proofs for the bounded commands.
By searching through the set of bounded proofs, \sys is searching for
synchronization points of the input programs that lead to a proof.
In \autoref{sec:choose}, we describe this generalization step in
detail.

If \sys cannot find a generalizable proof, then it increases $n$ and
starts again from the first step.

\subsubsection{Constructing Bounded Programs}
\label{sec:bounded}
In order to represent bounded versions of programs, we extend our
imperative command inductive definition with one additional
constructor, $\bot$.
$\bot$ should be understood as a command that immediately terminates.
We use this construction to replace
recursive calls to procedures outside the bound that we currently
consider.
$\Bounded$ is a procedure that constructs a bounded command $c'$
and corresponding lookup
table $T'$ from an input command $c$ with lookup table $T$ and a
bounding number $n$. 
The output command $c'$ is an under-approximation of
the input command $c$ that respects $n$.
The result of calling $\Bounded$ is a new, bounded program where each
recursive procedure is ``copied'' at most $n$ times. Further calls to
recursive procedures are modeled by $\bot$.
In \autoref{sec:overview}, \autoref{fig:bounded-code} shows a bounded
command $\cc{tri0}_0$ with its lookup table that constructs from
original command $\cc{tri0}$ in \autoref{fig:running-ex-code} with the
bounded number three. The missing `else' clauses in these examples
correspond to the command $\bot$.
An implementation of $\Bounded$ is described in
the extended paper.

\subsubsection{Verifying Bounded Programs via Constrained Horn Clauses}
\label{sec:solve}
%constructs a bounded command
\sys constructs invariants for all possible proofs of a bounded Hoare
Triple using a system of Constrained Horn Clauses (CHCs).
\autoref{alg:con-chc} describes $\ConstructCHC$, a procedure that
constructs a CHC system representing all possible proofs for a given
Hoare Triple.
The solution of a CHC system is a set of relational invariants that
support the set of proofs.
If \sys cannot find a solution of the constructed CHC system, then
either \textbf{(1)} \sys will find a counter-example of the relational
property or \textbf{(2)} the underlying theorem prover does not support
expressive enough logic to construct valid invariants.

\begin{algorithm}[t]
  \SetKwInOut{Input}{input}
  \SetKwInOut{Output}{output}
  \SetKwProg{myproc}{Procedure}{}{}
  \Input{%
    A Hoare Triple $\triple{P}{pcd'}{Q}$ where $pcd' \in \pcom$ is
    bounded product command and its lookup table $T$.
  }
  \Output{%
    A CHC system whose solution is a set of proofs for the given Hoare
    Triple.
  }
  \myproc{$\ConstructCHC(\triple{P}{pcd'}{Q},T)$} {%
    $CHC \gets \varnothing$\\
    \myproc{$\ConstructAux(\triple{P}{pcd'}{Q})$} {%
      \Switch{pcd'}{%
        \Case{$\cskip \times \cskip$}{%
          \Return{}
        }
        \Other{%
          \ForEach{$c_0 \times c_1 \in \Permute(pcd')$}{%
            \Switch{$c_0$}{%
              \Case{$\bot$}{%
                $CHC \gets CHC \cup \{ Q \Leftarrow \false \}$\\
              }\Case{$c_0';c_0''$}{%
                \ForEach{$(pcd_0, pcd_1) \in \Partition(pcd')$}{%
                  $R \gets \textsf{freshRel}$\\
                  $\ConstructAux(\triple{P}{pcd_0}{R})$\\
                  $\ConstructAux(\triple{R}{pcd_1}{Q})$\\
                }
              }\Case{$x := e$}{%
                $R \gets \textsf{freshRel}$\\
                $CHC \gets CHC \cup \{ [x \mapsto e]R \Leftarrow P \}$\\
                $\ConstructAux(\triple{R}{pcd'}{Q}$)
              }
              $\cdots$\\
            }
          }
        }
      }
    }
    $\ConstructAux(\triple{P}{pcd'}{Q})$\\
    \Return{$CHC$}
  }
  \caption{
    Given a Hoare Triple over a product command $pcd'$ and a
    corresponding lookup table $T$,
    generate a CHC system that represents all possible proof paths for
    this triple.
  }
  \label{alg:con-chc}
\end{algorithm}

$\ConstructCHC$ defines an auxiliary procedure $\ConstructAux$.
$\ConstructAux$ is a recursive descent over the structure of the
product command that accumulates a CHC system in the variable $CHC$.
If $pcd'$ is exactly the product command $\cskip \times \cskip$, then the recursion is finished
and $CHC$ contains a complete system.
Otherwise, $\ConstructAux$ applies the procedure $\Permute$ on the
product command $pcd'$.
$\Permute(pcd')$ returns two product commands by applying the proof rule
$\textsc{Comm}$.
For each $c_0 \times c_1$ in set $\Permute(pcd')$,
$\ConstructAux$ examines the first product $c_0$.
If $c$ is a sequence of commands, then $\ConstructAux$ applies
$\Partition$ to $c \times c_1$.
$\Partition$ is a procedure that generates a set of all possible
partitions of $pcd'$.
$\Partition(c_0)$ and $\Partition(c_1)$ are the sets containing all
valid partitions that respect the partition rule without duplicating $\cskip$.
For each pair $(pcd_0,pcd_1)$ in the set of partitions, $\ConstructAux$
constructs a fresh relational predicate $R$ as an intermediate
proposition. It then recurses on both parts.
If $c_0$ is neither a sequence of commands nor $\bot$, then
$\ConstructAux$ updates the CHC system based on the semantics of $c_0$
and recurses on $pcd'$. For example, if $c_0$ is an assignment then
a clause is added which indicates the precondition implies the
intermediate proposition with the appropriate substitution.

The key intuition behind $\ConstructCHC$ is that it constructs CHC
system that contains all possible proofs for the bounded command $pcd'$
by exhaustively applying the $\textsc{Comm}$ and $\textsc{Part}$ rules.
When the constructed CHC system is solved, the solution contains
invariants for all possible proofs of $\triple{P}{pcd'}{Q}$.
In practice, $\ConstructCHC$ includes optimizations that avoid redundant work.

\sys solves CHC systems generated by $\ConstructCHC$ using an external
solver.
The solution $\sigma$ is map from each relational predicate to its
interpretation. Replacing each relational predicate by the
corresponding invariant
in the proof leads to valid Hoare Triples for the bounded program.

\subsubsection{Generalizing Bounded Proofs}
\label{sec:choose}
\sys defines a procedure $\Synthesize$ that searches the set of proofs
for the bounded commands to find one proof that can be generalized for
the unbounded original commands.
$\Synthesize$ operates over a bounded product command $pcd'$ and
corresponding looks up table $T'$, as well as a solution $\sigma$ of
the constructed CHC system, which contains proper invariants for all
proof paths.
$\Synthesize$ decides if one generalizable proof path can be found for
the original, unbounded command $pcd$, within the current set of
bounded proof paths and its invariants.

The key intuition behind this algorithm is that $\Synthesize$ only
needs to find \textbf{one} generalizable proof path among the set of bounded proof 
paths with current synthesized invariants.
$\Synthesize$ has a similar structure to $\ConstructCHC$ with two key differences.

First, $\Synthesize$ attempts to use $\textsc{Assume}$ to generalize the current
proofs for the unbounded, original programs.
$\textsc{Assume}$ can only be applied when the context contains an
appropriate Hoare Triple as hypothesis.
$\Synthesize$ builds up context at each call site.
When it revisits an identical command a second time (called procedure
names can be different copies of the same original procedure),
$\Synthesize$ checks if the
pre-condition of the hypothesis is implied by the goal pre-condition
and if the post-condition of the hypothesis implies the goal
post-condition. 
If so, the hypothesis can be used to apply $\textsc{Assume}$ to find a generalized 
proof for the current goal.
For example, in \autoref{sec:example-proof},
\autoref{fig:subset-proof} depicts two bounded product commands
$\cc{tri0}_0 \times \cc{tri1Aux}_0$ and $\cc{tri0}_1 \times
\cc{tri1Aux}_1$. Both of these commands represent the same unbounded
product command, $\cc{tri0} \times \cc{tri1Aux}$.
Since the relational invariants for these two Hoare Triples are the
same, $\Synthesize$ can use the first triple as an assumption to prove
the second.

Second, $\Synthesize$ only needs to find one valid proof for the goal.
Thus $\Synthesize$ can choose between all permutations/partitions of
the bounded command $pcd'$ to find one generalized proof for the original command
$pcd$.
\autoref{fig:subset-proof} shows all possible proof paths for one
sub-proof goal, and $\Synthesize$ only needs to find one proof path that
passes through the node $\cc{tri0}_1 \times \cc{tri1Aux}_1$ to finish the
proof.
Other proof paths can be discarded.
The algorithm is presented more carefully in the extended paper.

\section{Evaluation}\label{sec:evaluation}
\begin{figure}
\centering
\begin{tabular}{l@{\hskip 0.27cm}
                l@{\hskip 0.27cm}
                l@{\hskip 0.27cm}
                r@{\hskip 0.27cm}
                r@{\hskip 0.27cm}
                l@{\hskip 0.27cm}
                l@{\hskip 0.27cm}
                }
\textbf{Source} &
\textbf{Name} &
\textbf{Property} &
\textbf{Time(s)} &
\textbf{Mem(MB)} &
\textbf{\cite{unno17}}&
\textbf{\cite{deangelis16}}
\\
\toprule
%Novel
%& $\textbf{triangle}$ &equiv & 5.8 & 168.8 & \xmark & \cmark\\
%\midrule
\multirow{15}*{\begin{tabular}{l}Automating\\ Induction\\ For
Solving\\ Horn Clauses \\ \cite{unno17}\end{tabular}}
& $\textbf{multMultAcc}$           & equiv & 5.6  & 183.0  & \cmark & \cmark\\
& $\textbf{multMultAcc0}$          & equiv & 5.5  & 182.8  & \cmark$^!$ & \cmark\\
& $\textbf{multL1}$                & equiv & 4.7  & 185.8  & \cmark & \xmark\\
& $\textbf{multR1}$                & equiv & 2.7  & 121.8  & \cmark & \cmark\\
& $\textbf{multDistL}$             & distr & 12.6 & 381.3  & \xmark & \xmark\\
& $\textbf{multDistR}$             & distr & 23.9 & 433.6  & \cmark & \xmark\\
& $\textbf{sumSimple}$             & equiv & 2.8  & 121.2  & \cmark & \cmark\\
& $\textbf{sumDown}$               & equiv & 5.2  & 197.1  & \cmark & \cmark\\
& $\textbf{sumUp}$                 & equiv & 5.0  & 177.3  & \cmark & \cmark\\
& $\textbf{sumUpDown}$             & equiv & 6.1  & 179.2  & \xmark & \xmark\\
& $\textbf{sumSumAcc}$             & equiv & 5.5  & 179.7  & \cmark & \cmark\\
& $\textbf{sumSumAcc0}^{\ddagger}$ & equiv & 11.7 & 253.6  & \xmark & \cmark\\
& $\textbf{multAssoc}$             & assoc & 33.3 & 599.4  & \xmark & \xmark\\
& $\textbf{sumMono}$               & mono & TO  & 1530.6 & \cmark & \cmark\\
& $\textbf{multMono}$              & mono  & TO  & 1368.3 & \cmark$^!$ & \cmark\\
& $\textbf{multComm}$              & comm  & TO  & 2028.8 & \cmark & \cmark\\
\midrule
\multirow{10}*{\begin{tabular}{l}Software\\ Foundations\end{tabular}}
& $\textbf{plusComm}$             & comm  & 6.0   & 254.2 & \xmark & \cmark\\
& $\textbf{plusAssoc}$            & assoc & 38.9  & 618.0 & \xmark & \xmark\\
& $\textbf{plusNSm}$              & equiv & 21.0  & 430.6 & \cmark & \cmark\\
& $\textbf{plusNSm0}^{\ddagger} $ & equiv & 22.9  & 434.7 & \xmark & \cmark\\
& $\textbf{plusRearrange}$        & equiv & 138.5 & 658.6 & \xmark & \cmark\\
& $\textbf{doublePlus}$           & equiv & 4.3   & 118.2 & \xmark & \cmark\\
& $\textbf{doubleInjective}$      & inj   & 4.8   & 237.5 & \xmark & \cmark\\
& $\textbf{evenbS}$               & equiv & 26.6  & 642.2 & \xmark & \xmark\\
& $\textbf{beqNatSym}$            & sym   & 5.4   & 202.3 & \cmark & \cmark\\
& $\textbf{beqNatTrans}$          & tran  & 7.2   & 242.4 & \cmark & \cmark\\
& $\textbf{mult0plus}$            & equiv & TO   & 786.1 & \xmark & \xmark\\
\midrule
\multirow{5}{*}{LeetCode}
& $\textbf{addDigits}^{\dagger}$   & equiv & 2.5 & 67.2 & \xmark & \cmark\\
& $\textbf{trailingZeroes}$        & equiv & 5.1 & 200.7 & \xmark & \cmark\\
& $\textbf{climbStairs}^{\dagger}$ & equiv & 6.3 & 258.0 & \xmark & \xmark\\
\end{tabular}

\caption{The results of our evaluation of \sys.
  Each benchmark is labeled with its source, name, the class of
  relational property that \sys attempted to verify, time spent by
  \sys to synthesize a proof, the peak amount of memory that \sys
  used, and whether \emph{automated induction} \cite{unno17} or
  \emph{VeriMapRel}\cite{deangelis16} verified the benchmark.
  A time of \emph{TO} denotes that \sys was unable to converge within 
  $300$ seconds.
  The superscript `!' denotes that automated induction only converged
  with a manually-provided lemma.
  Each benchmark with the superscript `$\ddagger$' is a minor
  modification of the original benchmark immediately above it.
  The superscript `$\dagger$' denotes that the benchmarks obtained
  from the source were not equivalent.
  In such cases, the data reports the performance of \sys when applied
  to a version of the benchmark that we manually patched to be
  correct.  }
\label{fig:results}
\end{figure}

We performed an empirical evaluation of \sys to answer the following
questions:
How effective is \sys compared to other automated
relational verifiers?

% summary of implementation, benchmarks:
To answer the above experimental questions, we implemented \sys as a
verifier of relational properties of programs represented in JVM
bytecode.
The only requirement imposed by \sys~on the logic for expressing
program semantics is that the logic %
\textbf{(1)} has an effective decision procedure, which \sys~uses to
check possible entailments (\autoref{sec:solve}), and %
\textbf{(2)} can be encoded in the logic of constraints supported by
its CHC solver.
A subset of the JVM semantics can be encoded in the logic of linear
arithmetic with arrays. This logic is supported both by the Z3 decision
procedure and the \textsc{Duality} CHC solver implemented in
Z3~\cite{z3}.

% describe benchmarks:
We applied \sys~to benchmarks introduced in previous work on
relational verification by \emph{automatic induction}~\cite{unno17},
programs and properties corresponding to theorems over recursive
functions posed as theorem-proving exercises\cite{pierce17}, and
solutions to coding problems on the Leetcode platform~\cite{leetcode}.
We also slightly modified two benchmarks (\cc{plusNSm0} and
\cc{sumSumAcc0}) that required a verifier to prove a corollary that is
strictly weaker than key inductive mutual summary of the programs.
Such modified benchmarks can present distinct challenges to a verifier
because they require the verifier to synthesize non-trivial inductive
summaries.
The benchmarks require proofs of properties including equivalence,
distributivity, monotonicity, commutativity, associativity,
injectivity, transitivity, and symmetry.

% describe baseline:
We compared \sys~to implementations of techniques that perform
automatic induction~\cite{unno17}, that transform CHC systems encoding
relational properties (VeriMapRel)~\cite{deangelis16}, that use
Cartesian Hoare Logic (CHL)~\cite{sousa16}, and that use
self-composition.
The current implementation of CHL does not support recursive
procedures and self-composition cannot solve any but the simplest
problem, \cc{addDigits}.
VeriMapRel does not support the negation of equality statements in its
property specification, so
we have to manually transformed the benchmarks with equality
statements to a set of relational properties that use inequalities.
Without this manual work, VeriMapRel can only solve two benchmarks.
As a result, we have reported comparisons with automatic
induction, and with VeriMapRel using this manual transformation.

% walk through the results:
\autoref{fig:results} contains the results of our evaluation.
In short, our experiments indicate that \sys can efficiently verify
properties beyond the scope of existing techniques.
In particular, \sys successfully verifies all but four of the
benchmarks on which it was evaluated.
Automatic induction fails to prove 15 cases within time that \sys can.
These cases require
synthesizing non-trivial inductive relational invariants other than
the given relational properteis to finish the proof.
VeriMapRel fails to prove 8 cases within time that \sys can.
These cases requires sophisticated synchronization between two programs.

\sys failed to converge on four cases because \textsc{Duality} did
not generate relational invariants of bounded programs that can be generalized.
This is a known challenge for CHC solvers that use an interpolating
theorem prover~\cite{albarg13}.
For example, to prove that multiplication is commutative
(\textbf{multComm}), \sys requires the CHC solver to generate
summaries that establish equalities over program variables, such as
$\cc{x}_0 = \cc{y}_1$.
Instead, the solver sometimes generates invariants specific to the
structure of the hierarchical programs, such as
$\cc{x}_0 = 1 \land \cc{y}_1 = 1$.
However, because \sys uses a CHC solver as a black box, it is well
positioned to benefit directly from improvements to CHC solvers.
Furthermore, in a significant number of cases, \sys synthesize proper synchronization
points with relational invariants from \textsc{Duality}'s solutions that could
not be found by existing techniques.
The current implementation of Pequod and executable benchmarks are
available
online.\footnote{https://www.dropbox.com/s/yks0eyic8dsf69e/pequod.zip?dl=0}

\section{Related Work}\label{sec:related-work}

% automating induction:
Previous work~\cite{deangelis16,kiefer16,unno17} has established that
problems in relational verification can be reduced to solving systems
of Constrained Horn Clauses, and has proposed novel proof systems for
solving such systems.
Such systems are expressive, and can be partially automated.
However, they require a prover to manually provide lemmas that the
system establishes by induction when a lemma stronger than the goal
invariant must be proved~\cite{unno17} (analogous to suggesting
inductive invariants when they must differ from the goal invariant of
a program) or direct how relational predicates in a given system
should be paired in order to generate a solvable
system~\cite{deangelis16,kiefer16}.
\sys~performs such reasoning automatically.

% other relational proof system
Previous work has proposed frameworks that allow a prover to verify
that recursive programs satisfy a mutual
summary~\cite{banerjee16,backes13,bohme13,godlin09,hawblitzel13}, but
require the user to direct how procedures must be paired, and in some
cases provide mutual summaries.
Other approaches for verifying relational properties of
single-procedure programs have been significantly
automated~\cite{sousa16}, but the developed automation tactics are
carefully tuned to syntactic forms of the programs and would be
non-trivial to generalize to programs that contain multiple
procedures.

% constructing product programs:
Verifying relational properties can also be reduced to synthesizing a
suitable \emph{product program}~\cite{barthe11,barthe16}.
Some approaches synthesize product programs in the class of
\emph{sequential compositions} automatically, but such product
compositions either cannot easily be constructed
manually~\cite{beringer11} or can only prove relational properties in
a heavily restricted
class~\cite{barthe04,barthe11,felsing14,lopes16,terauchi05}.
Other approaches construct product programs depending partly on
matching control structures between the pairs of programs and
establishing the logical equivalence of program conditions included in
matched structures.
% us: more restricted, but automatic
Previous work has also explored constructing \emph{asymmetric product
  programs}~\cite{barthe13} which can express proofs of relational
properties not provable in the system used by \sys.
However, such work does not address the problem of automatically
inferring loop invariants of the synthesized product program, which
may be viewed alternatively as the mutual summary between loops of the
original programs.

% logic for higher-order programs:
Recent work has introduced logics for reasoning about relational
properties of higher-order programs~\cite{aguirre17}.
However, these systems have not yet been used to automatically
synthesize proofs of program equivalence.
\sys~can only synthesize proofs for first-order recursive programs,
but can do so automatically.

\bibliographystyle{eptcs}
\bibliography{p,conf}

\end{document}